\documentclass{iau} 
\usepackage{graphicx}

\title[The HERON Survey] 
{The Halos and Environments of Nearby Galaxies (HERON) Survey}

\author[R. Michael Rich]   
{R. Michael Rich $^1$, Noah Brosch $^2$, James Bullock $^3$, Andreas Burkert$^4$, Michelle Collins$^5$, Laura de Groot$^6$, Julia Kennefick $^7$, Andreas Koch $^8$, Francis Longstaff $^9$,
 \and Laura Sales $^{10}$}

\affiliation{$^1$ Dept. of Physics and Astronomy, UCLA, Los Angeles, CA 90095-1547 \\ email: {\tt rmr@astro.ucla.edu}\\ [\affilskip] $^2$ Wise Observatory, Tel Aviv University $^3$ UC Irvine $^4$ Ludwig- Maximillians Universitat Munchen $^5$ University of Surrey $^6$ Denison University, $^7$ Uinv. of Arkansas $^8$ Univ. of Lancaster $^9$ Anderson School of Management, UCLA $^{10}$ UC Riverside}

\pubyear{2016}
\volume{321}  
\jname{Outskirts of Galaxies}
\editors{A.C. Editor, B.D. Editor \& C.E. Editor, eds.}
\begin{document}

\maketitle

\begin{abstract}
We have used dedicated 0.7m telescopes in California and Israel to image the halos of $\sim 200$ galaxies in the Local Volume to 29 mag/sq arcsec, the sample mainly drawn from the 2MASS Large Galaxy Atlas (LGA).  We supplement the LGA sample with dwarf galaxies and more distant giant ellipticals. Low surface brightness halos exceeding 50 kpc in diameter are found only in galaxies more luminous than $L^*$, and classic interaction signatures are relatively infrequent. Halo diameter is correlated with total galaxy luminosity. Extended low surface brightness halos are present even in galaxies as faint as $M_V=-18$. Edge-on galaxies with boxy bulges tend to lack extended spheroidal halos, while those with large classical bulges exhibit extended round halos, supporting the notions that boxy or barlike bulges originate from disks. Most face-on spiral galaxies present features that appear to be irregular extensions of spiral arms, although rare cases show smooth boundaries with no sign of star formation.  Although we serendipitously discovered a dwarf galaxy undergoing tidal disruption in the halo of NGC 4449, we found no comparable examples in our general survey.    A search for similar examples in the Local Volume identified hcc087, a tidally disrupting dwarf galaxy in the Hercules Cluster, but we do not confirm an anomalously large half-light radius reported for the dwarf VCC 1661.
\end{abstract}

\begin{figure}
\begin{center}
 \includegraphics[width=6.5 in]{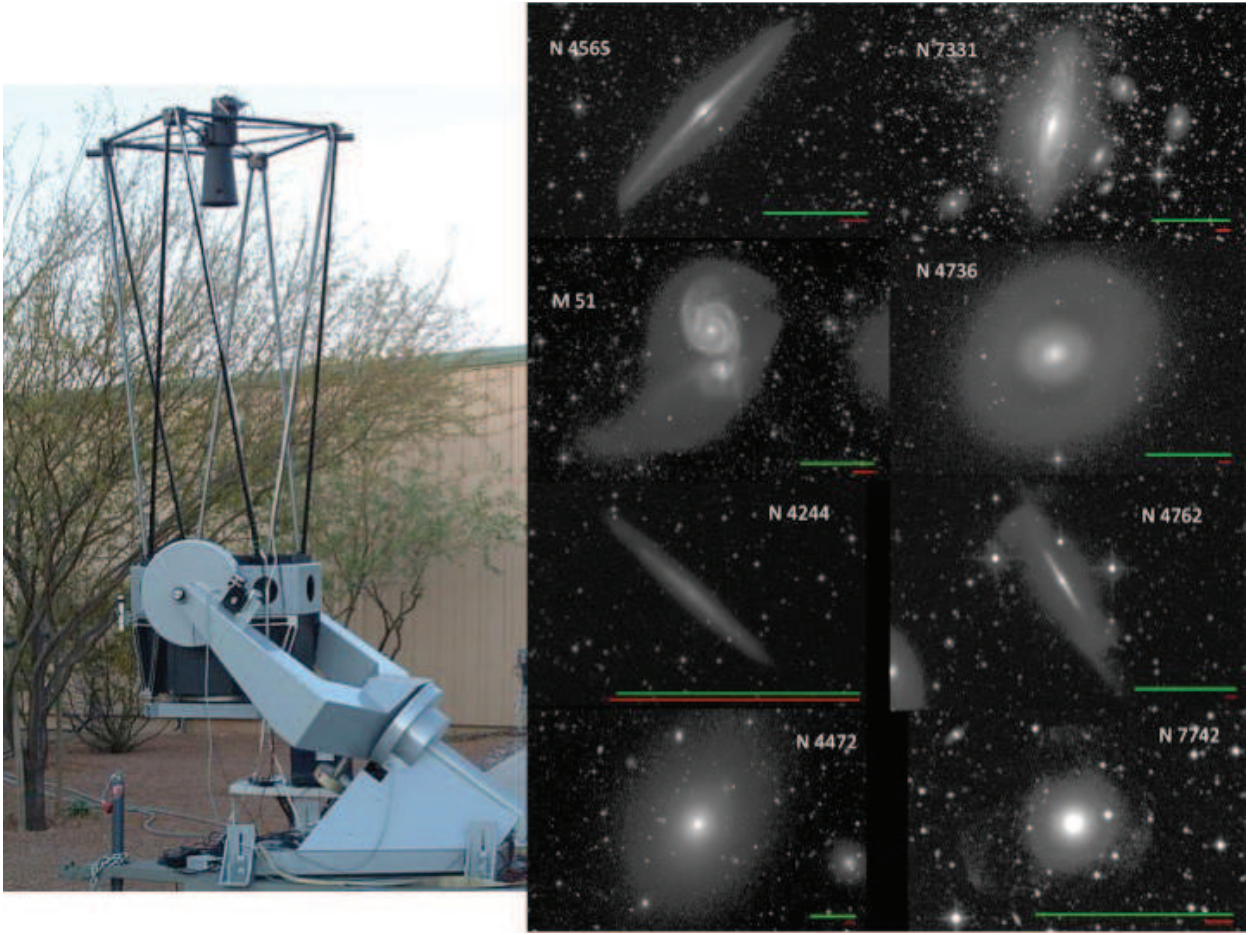} 
 \caption{(Left) the 28-inch=0.7m telescope during final assembly near Aguila Arizona. (Right) Recent results obtained with our facility.  The green (upper) scale bar is 5'; the red (lower) scale bar is 10 kpc at the distance of the galaxy as obtained from the National Extragalactic Database (NED).  These images show features to roughly 29 mag/sq arcsec and most exposures are 300 sec subexposures totalling 1 hour.  The halos show an order of magnitude range in physical sizes.}
   \label{fig1}
\end{center}
\end{figure}

Very deep imaging of nearby galaxies is not a recent venture.  The work of David Malin using the technique of unsharp masking applied to photographic plate images was both pioneering and demonstrated the potential science return from low surface brightness imaging (Malin 1978; Malin \& Carter 1980).  The recent advent of relatively low cost CCDs and the plethora of telescopes of semi-professional quality at $\sim 0.5$m aperture has resulted in a flood of observations of interesting structures and streams; see e.g. \cite{md10}.  The field now includes effort that aggregate long exposures taken by amateur astronomers and new larger dedicated efforts like the {\sl Dragonfly} project; \cite{dragonfly}.   The {\sl HERON} project has a somewhat unusual history.  Francis Longstaff, a professor of finance at UCLA, was interested in astronomical imaging and had collaborated on data analysis with the UCLA Galactic Center group.  He initiated the purchase of the 28-inch (0.7m) Centurion telescope designed by J. Riffle, Astrowork, Inc., which built roughly 100 18-inch telescopes but only 3 of the 28-inch class; Rich and Longstaff collaborated on the acquisition and installation of the facility.  The telescope is shown in Fig. \ref{fig1}, and is described further in \cite{brosch15}.   In contrast to most telescopes developed for the consumer market, the C28 places its detector at the f/3.2 prime focus behind a 2-element Ross corrector that includes a conical baffle; the primary mirror is hyperbolic and the mount is an equatorial yoke.   This design renders a fast system that can deliver images as good as $1.5$ arcsec.  The site is near Frazier Park, CA at an altitude of 1583 m; it has been an amateur astronomy club retaining dark skies for a generation, being founded in 1966.  While the light dome of Los Angeles has grown over the years, the zenith on moonless nights is typically 21.65 mag/sq arcsec with the West being very dark.   Early success using the telescope for low surface brightness imaging (Rich et al. 2012) inspired efforts to raise funds to erect a sister telescope of the same aperture at the Wise Observatory  in Israel.   Initial operations employed at the California site began with the SBIG STL11000 imager which has been upgraded to an FLI09000 imager presently on loan from Arizona State University.  Total operations costs are $<\$5,000$/yr.

\begin{figure}
\begin{center}
 \includegraphics[width=4.5 in]{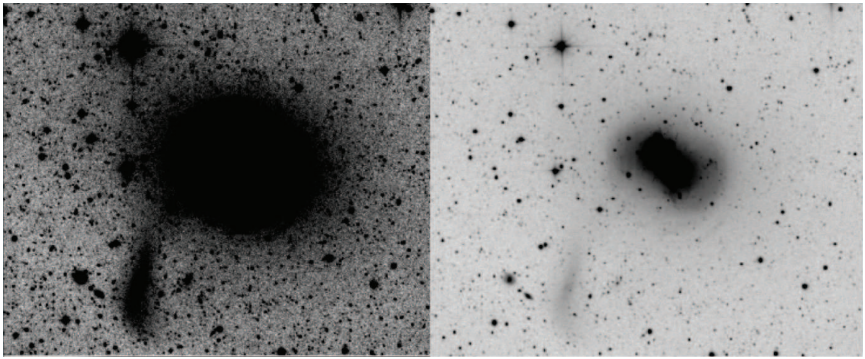} 
 \caption{(Left) Hard stretch of NGC 4449 showing the outer halo and NGC 4449B, the dwarf galaxy (Rich et al. 2012).  (Right) NGC 4449 showing the m=2 distortion and ansae in the disk portion just outside of the central bar; the surface brightness profile follows an exponential disk.}
   \label{fig2}
\end{center}
\end{figure}

\begin{figure}
\begin{center}
 \includegraphics[width=7.0 in]{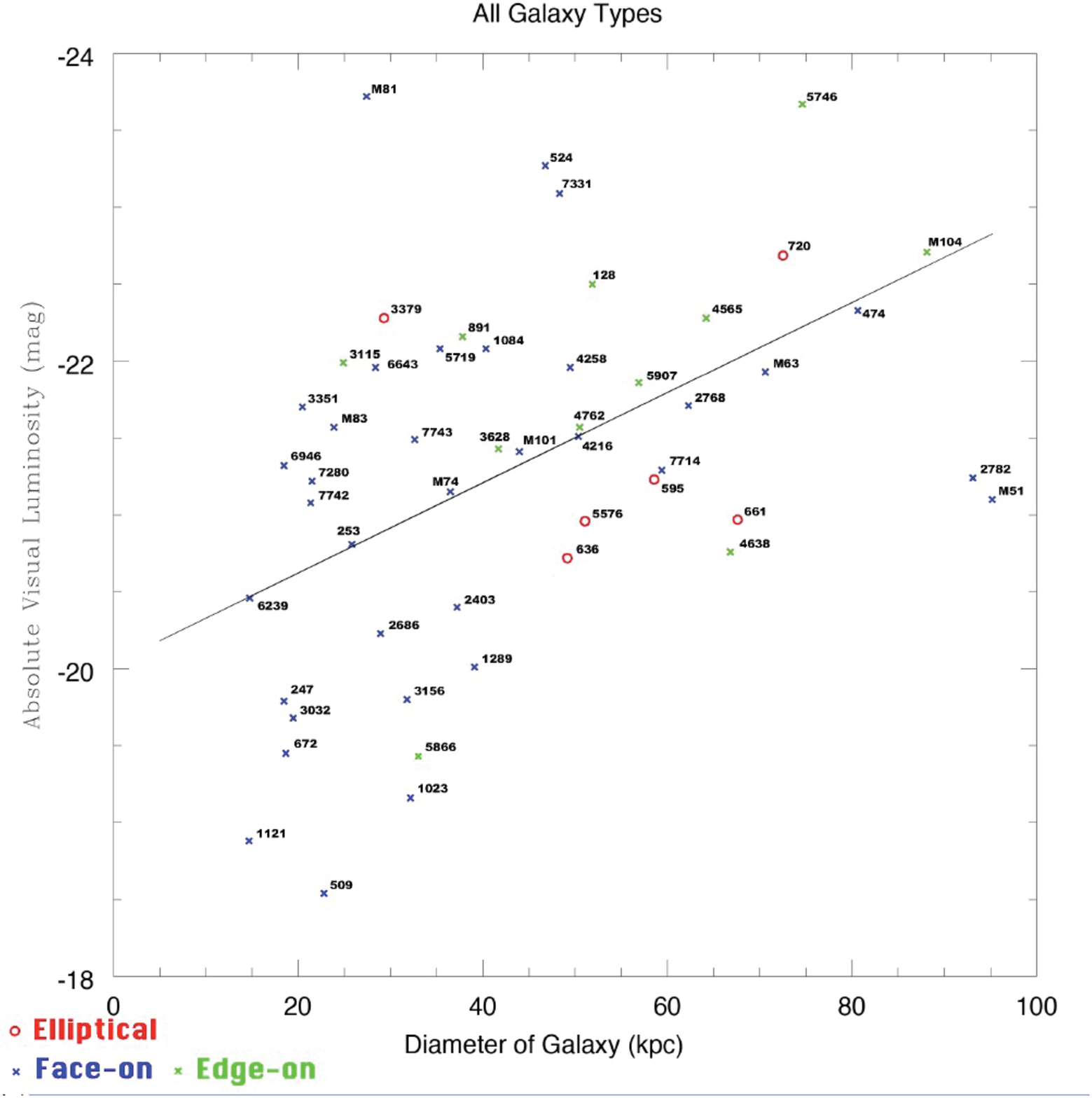} 
 \vspace*{-0.2 cm}
 \caption{ Absolute magnitude of the host as a function of total halo diameter (including measurable streams) at 29 mag $\square " ^{-2} $.  Note that only galaxies $M_V<-21$ possess "giant" halos larger than 40 kpc.  Galaxies with $M_V<-21$ also are known to have "metal rich" stellar halos with $\rm [Fe/H]\ > -0.7$ (Mouhcine et al. 2005) }.
   \label{fig3}
\end{center}
\end{figure}

Our survey is complete for $\delta > -20^\circ $ in the 2MASS large galaxy atlas (Jarrett et al. 2003); we supplement our survey with some nearby low luminosity systems.  One case is NGC 4449, which resulted in the discovery the tidal dwarf NGC 4449B (Fig.\,\ref{fig2}).  This object clearly lies above the trend line of $r_h \rm{vs}\ M_v$ (Misgeld \& Hilker 2011), as does hcc087 in the Hercules cluster.  Clearly well above the trend line, hcc087 is confirmed by \cite[Koch et al. (2012)]{koch12} as a striking tidally disrupting dwarf with tidal tails.    In contrast, the dwarf galaxy VCC1661 was reported by \cite[Ferrarese et al. (2006)]{ferr06} to have half-light radius $r_e=4.63\pm 0.33$ kpc; the position on the $r_h$ - $M_v$ plot marked it as a candidate to be an especially extended tidally disrupting dwarf.  Our observations found it instead to be normal, with $r_e=1.93 \pm 0.63$ kpc (Koch et al. 2016).  Indeed, our survey found no new tidally disrupting dwarfs.  We have found a general trend that halos larger than 40 kpc are hosted only in galaxies with $M_V<-21.5$ (Fig.\,\ref{fig3}).  Edge-on boxy bulges such as NGC 5746 lack a round halo, with their lowest surface brightness extent being a flattened ellipse and aligned with the disk, while NGC 7331 and other galaxies with more prominent bulges are dominated by round isophotes that extend to the faintest levels.
 We are preparing a comprehensive report on our survey for publication later this year.

\vspace*{-0.2 cm}

\end{document}